\documentclass{moriond}

\def\be{\begin{equation}}
\def\ee{\end{equation}}
\def\bea{\begin{eqnarray}}
\def\eea{\end{eqnarray}}

\newcommand{\Photo}{}

\begin{document}
\vspace*{4cm}
\title{Constraints on the positron emission from pulsar populations with AMS-02 data}

\author{ Silvia Manconi }

\address{Institute for Theoretical Particle Physics and Cosmology, RWTH Aachen University\\ Sommerfeldstr.\ 16, 52056 Aachen, Germany}

\maketitle\abstracts{
Electron and positron fluxes in cosmic rays are currently measured with unprecedented precision by AMS-02 up to TeV energies, and represent unique probes for the local properties of our Galaxy. 
The interpretation of their spectra is at present still debated, especially for the excess of positrons above 10 GeV. 
The hypothesis that pulsars can significantly contribute to this excess has been consolidated after the observation of gamma-ray halos at TeV energies of a few degree size around Geminga and Monogem pulsars.
However, the spatial and energetic Galactic distribution of pulsars and the details of the positron production, acceleration and release from these sources are not yet fully understood.
I will describe how we can use the high-precision AMS-02 positron data to constrain the main properties of the Galactic pulsar population and of the positron acceleration needed to explain the observed fluxes. This is achieved by simulating a large number of Galactic pulsar populations, following the most recent self-consistent modelings for the pulsar spin-down and evolution properties, calibrated on catalog observations. By fitting the positron AMS-02 data together with a secondary component due to collisions of primary cosmic rays with the interstellar medium, we determine  the physical parameters of the pulsars dominating the positron flux, and assess the impact of different assumptions on radial distributions, spin-down properties, Galactic propagation scenarios and positron emission time.}

\section{Introduction}
\textit{This contribution is based on this paper \cite{Orusa:2021tts} done in collaboration with Luca Orusa, Mattia Di Mauro, Fiorenza Donato (University of Turin and INFN, Turin), and on additional reasoning by the speaker, following the material presented during the talk at the conference. I refer to the paper for an extended discussion of the model, the results, and for a more exhaustive bibliography, reduced to minimal here due to space limitations.}

\subsection{The local $e^+$ flux}
High-energetic cosmic-ray (CR) positrons ($e^+$) contribute to the total CR flux detected at Earth at the percent level. Nevertheless, the unprecedented precision of the measurement of the local $e^+$ flux by PAMELA,  Fermi-LAT and most recently by AMS-02 \cite{PhysRevLett.122.041102} allowed an in-depth study of this antimatter component of CRs. This suggests the presence of primary $e^+$ sources in our Galaxy.
In fact, the observed flux exceeds the so-called secondary flux (as computed in the standard CR transport models, see later) produced by inelastic collisions of CR nuclei in the interstellar medium (ISM) above about 10~GeV. 
For an overview on the most recent CR observations and on the prospects for current and future observatories, we refer to other contributions to this conference \cite{Vecchi}.

The quest for the interpretation of the $e^+$ flux has been a dynamic part of the CR community's debate since more than 10 years. 
The intense radiative losses suffered by high energetic $e^+$ while propagating in the Galaxy limit the horizon for GeV-TeV $e^+$ to few kiloparsecs (kpc). 
Thus, it is generally accepted that AMS-02 $e^+$ observations require their primary source to be local, i.e.~located within a few kpc from the Earth.
Many mechanisms have been explored in the literature to account for the  $e^+$ flux observed by AMS-02. Among them, pulsars and their pulsar wind nebulae (PWNe) have been consolidating as significant factories of high-energy CR $e^\pm$ in the Galaxy, and thus as main candidates to explain the $e^+$ excess.
A not-exhaustive list of the main alternative proposals  includes dark matter annihilations in the Galactic halo \cite{2016JCAP...05..031D}, modifications of the standard secondary production mechanism \cite{Lipari:2018usj}, as well as production of secondary antimatter in supernova remnants,  also discussed at this conference \cite{Philipp}. 
  
\subsection{Pulsars and their nebulae as $e^+$ factories}

Many theoretical and observational arguments support the idea that pulsars and their nebulae are $e^+$ (and electrons  \footnote{Galactic
pulsars are expected to contribute to the data on the $e^+ + e^-$ and electron ($e^-$) spectrum as well, see e.g.\cite{2019_Manconi,Evoli_2021}. 
 However, the AMS-02 $e^+$ data up to 1~TeV are currently the most precise observable to constrain the characteristics of Galactic pulsar populations, being this the main scope of the present work. 
The $e^+ + e^-$ data is indeed dominated by the $e^-$ produced by supernova remnants (SNR)\cite{2019_Manconi,Evoli_2021}.}) factories. 

First, the pulsar spin-down mechanism  effectively 
produces $e^\pm$ pairs, which are possibly accelerated to multi-TeV energies at the termination shock between the relativistic wind and the surrounding medium (see \cite{Amato:2020zfv} for recent reviews on the acceleration mechanism).
We note that the details of the $e^\pm$ production, acceleration and release from pulsars and their PWNe are yet not fully understood \cite{Amato:2020zfv}, as well as  the spatial and energetic distribution of pulsar \cite{Lorimer_2006,Chakraborty:2020lbu}. 
CR $e^+$  are thus a golden channel to study  primary sources, and in particular PWNe.

The observation of gamma-ray halos at TeV energies of a few degree size around nearby pulsars \cite{Mattia}, such as the ones of Geminga and Monogem, reported by HAWC  \cite{Abeysekara:2017science}  and 
by {\it Fermi}-LAT at tens of GeV \cite{DiMauro:2019yvh}
further corroborates the presence of $e^+$ (and electrons) accelerated, then escaped at few tens of parsec away from the pulsar location. 
The observed emission is interpreted as coming from CR $e^\pm$ escaping from the  PWN  and inverse Compton scattering (ICS) low-energy photons of the interstellar radiation field (ISRF), see other contributions  to this conference for an extended discussion \cite{Mattia}. 

 Finally, several independent works have demonstrated that pulsar models can provide a good description of AMS-02 $e^+$  and $e^-$  data. 
This conclusion has been reached both by considering the contribution of few nearby sources, as well as the cumulative emission from pulsars as observed in existing catalogs (e.g. \cite{Fornieri_2020,DiMauro:2020cbn} and references therein) or in simulations \cite{Cholis_2018,Evoli_2021}, and 
including or not possible effects from suppressed diffusion around sources \cite{Manconi:2020ipm}, as suggested by the measured angular profile of gamma-ray pulsar halos \cite{Mattia}.
While energy losses limit the distance traveled by  high-energy $e^\pm$ to few hundreds of parsecs, where we expect few Galactic sources contributing significantly, current source catalogs might be not complete. Previous computations calculating the contribution of $e^+$ from the ATNF catalog sources 
could therefore suffer for underestimation due to the incompleteness of the catalog. Simulations of the Galactic source population of pulsars are thus needed to extensively test the pulsar interpretation of the observed $e^+$ flux in order to overcome the limitations of previous studies, see also \cite{Mertsch:2018bqd}.
Following the arguments introduced until now, high-precision $e^+$ data can now be used to constrain the main properties of the Galactic pulsar population and of the PWN acceleration. 
In this contribution  we build upon previous works \cite{Cholis_2018,Manconi:2020ipm,Evoli_2021} and extend them significantly in various novel aspects \cite{Orusa:2021tts}. 

\section{Phenomenological models for pulsar populations}

We simulate Galactic pulsars following the injection and propagation model summarized below, and produce mock catalogs for five simulation setups (\texttt{ModA-B-C-D-E}). 
For each realization we compute the $e^+$ flux from every PWN. 
A summary of the simulated quantities is illustrated in Table~1 in \cite{Orusa:2021tts} and outlined in what follows.

\subsection{Pulsar $e^+$ injection}
Pulsars are rotating neutron stars with a strong surface magnetic field, and magnetic dipole radiation is believed to provide a good description for its observed loss of rotational energy \cite{Amato:2020zfv}.
We consider a model in which $e^{\pm}$ are continuously injected at a rate that follows the pulsar spin-down energy. 
The injection spectrum $Q(E,t)$ of $e^\pm$ at energy $E$ and time $t$ is described as:
\begin{equation}
\label{eq:spectrum}
    Q(E,t)=L(t) \left(\frac{E}{E_0} \right)^{-\gamma_e} \exp \left(- \frac{E}{E_c} \right) 
\end{equation}
where the cut-off energy $E_c$ is fixed at $10^5$ TeV, $E_0 = 1$  GeV and $\gamma_e$ is the $e^\pm$ spectral index. The magnetic dipole braking $L(t)$ is described by the function $L(t)={L_0}/{\left( 1+ \frac{t}{\tau_0}\right)^{\frac{n+1}{n-1}}}$,
where $\tau_0$ is the characteristic time scale and $n$ defines the magnetic braking index.
The total energy emitted by the source only into $e^+$ is given by $E_{tot}=\eta W_0= \int_{0}^T dt \int_{E_1}^{\infty} dE E Q(E,t)$,
through which we obtain the value of $L_0$, fixing $E_1$=0.1 GeV. The parameter $\eta$ encodes the efficiency of conversion of the spin-down energy into $e^+$(which is half of the efficiency of conversion into $e^\pm$).
$W_0$ is the initial rotational energy of a pulsar with a moment of inertia $I$ and rotational frequency $\Omega_0=2\pi/P_0$, $W_0=E_{\rm rot,0}=\frac{1}{2} I {\Omega_{0}}^2\,,$
and it's determined for each mock pulsar using the pulsar spin down model below. 
Since the spectral index $\gamma_e$ of accelerated particles is uncertain, and may vary significantly for each  PWN, it is sampled from uniform distributions within [1.4-2.2]. The value of $\eta$ for each source is sampled from a uniform distribution in the range [0.01-0.1].

\subsection{Pulsar spin down}
In each simulation, the total number of sources is fixed at $N_{\rm PSR} = t_{max} \dot{N}_{PSR}$, where $t_{max}$ is the maximum simulated age and $\dot{N}_{PSR}$ is the pulsar birth rate.  We here assume the maximum age of the sources to be $t_{max}$ = $10^8$ yr, and $\dot{N}_{PSR}$ = 0.01 yr$^{-1}$.

The spin-down luminosity $\dot{E} = dE_{\rm rot}/dt$ of a pulsar is the rate at which the rotational kinetic energy is dissipated, and it's related to the pulsar period and period derivative as $\dot{E}=\frac{d E_{\rm rot}}{dt}= I \Omega \dot{\Omega}= -4\pi^2 I \frac{\dot{P}}{P^3}\,. $
Assuming a small deviation from the dipole nature of the magnetic field  $B$ of the pulsar, the evolution of the star may be parameterized as $P^{n-2}\dot{P} = a k (B \sin \alpha)^2\,.$ \cite{Ridley_2010},
where the angle $\alpha>0$ describes the inclination of the magnetic dipole with respect to the rotation axis, $a$ is a constant of unit s$^{n-3}$ and $k$ takes the value of $9.76 \times 10^{-40}$ s\,G$^{-2}$ for canonical characteristics of neutron stars. 
Within this model, the spin-down luminosity evolves with time $t$ as $ \dot{E}(t) = \dot{E}_0 \left( 1+ \frac{t}{\tau_0}\right)^{-\frac{n+1}{n-1}}.$
Finally, 
the prediction on the typical decay time $\tau_0$ is derived to be $\tau_0= \frac{P_0}{(n-1)\dot{P}_0}.$
By extending the functions implemented in the Python module {\tt gammapy.astro.population} \cite{Nigro_2019,CTAConsortium:2017xaq}, we sample the values of $P_0, B, n$ and $\alpha$ from the distributions provided in \cite{Chakraborty:2020lbu} (\texttt{CB20}), which will be our benchmark model, see Table~1 in \cite{Orusa:2021tts} for the parameter intervals.

In order to assess the effect of different distributions for $P_0, B, n$ and $\alpha$, we  consider the model in \cite{2006ApJ...643..332F} (\texttt{FK06} hereafter).  
We note that both \texttt{CB20} and \texttt{FK06} models have been calibrated to reproduce the characteristics of the sources detected in the ATNF catalog \cite{2005AJ....129.1993M}, like the $P$, $\dot{P}$, $B$, flux densities at 1.4 GHz, Galactic longitudes and Galactic latitudes distributions. \texttt{CB20} is the most updated model and considers the variation of more parameters with respect to \texttt{FK06}.

\subsection{Release, and propagation around sources}
In our benchmark model we will consider only sources with ages above 20 kyr, since $e^\pm$ accelerated to TeV energies in the termination shock are believed to be confined in the nebula or in the SNR until the merge of this system with the ISM, estimated to occur some kyr after the pulsar formation. We thus leave out sources for which the $e^\pm$  pairs might be still confined in the parent remnant. However, this effective treatment does not account for possible spectral or time-dependent modifications of the released particles. To understand the consequences of this assumption on the interpretation of the AMS-02 $e^+$ flux, we also test the hypothesis that only the $e^\pm$ produced after the escaping of the pulsar from the SNR contribute to the flux at the Earth. Following \cite{Evoli_2021}, we define $t_{BS}$ as the time at which the source leaves the parent SNR due to its proper motion and eventually forms a bow-shock nebula.  The formalism is reported in ~\cite{van_der_Swaluw_2003}, to which we refer for further details. 
To test this scenario we additionally simulate for each source its birth-kick velocity, adopting its distribution as reported in \cite{2006ApJ...643..332F} (\texttt{FK06VB}) and implemented in  {\tt gammapy.astro.population} \cite{Nigro_2019,CTAConsortium:2017xaq}.

\subsection{Pulsar spatial distribution}
Using {\tt gammapy.astro.population} \cite{Nigro_2019,CTAConsortium:2017xaq} we adopt the radial surface density of pulsars $\rho_L(r)$ proposed by \cite{Lorimer_2006}. 
As a comparison, we also consider the radial surface density $\rho_F(r)$  in \cite{2006ApJ...643..332F}. We sample the position ${\bf r}$ of each source combining the radial surface density with the  spiral arm structure of the Milky Way of ~\cite{2006ApJ...643..332F} (see their Table 2 for the spiral arm parameters), 
as implemented in {\tt gammapy.astro.population} \cite{Nigro_2019,CTAConsortium:2017xaq}. We test only one spiral arm structure, since the most important aspect in the computation of the $e^+$ flux is the source density in the arms nearby the Sun, instead of the position of the arms themselves.
The distance of each source is $d$=$|{\bf r}-{\bf r}_{\odot}|$, with ${\bf r}_{\odot} = (8.5, 0, 0) {\rm kpc}$.

\section{Transport in the Galaxy}
Once charged particles are injected in the Galaxy, they can propagate and eventually reach the Earth. 
The  number density per unit time, volume $N_e(E, {\bf r}, t)$ of $e^{\pm}$ at an observed energy $E$, a position ${\bf r}$ in the Galaxy, and time $t$, which is the solution to the propagation equation considering only diffusion and energy losses, is given by \cite{DiMauro:2019yvh}:
\begin{equation}\label{eq:numberdensity}
    N_e(E, {\bf r}, t)=\int_0^t dt' \frac{b(E_s)}{b(E)} \frac{1}{(\pi \lambda^2(t',t,E))^{\frac{3}{2}}}  \exp \left(-\frac{| {\bf r}-{\bf r_s}|^2}{\lambda ^2 (t',t,E)} \right) Q(E_s,t')
\end{equation}
where the integration over $t'$ accounts for the PWN releasing $e^{\pm}$ continuously in time. The energy $E_s$ is the initial energy of $e^{\pm}$ that cool down to $E$ in a loss time $ \Delta \tau \equiv \int_E^{E_s} \frac{dE'}{b(E')} = t - t_{obs}.$
The $b(E)$ term is the energy loss rate, ${\bf r_s}$ indicates
the source position, and $\lambda$ is the typical propagation length defined as $ \lambda^2 = \lambda^2(E,E_s) \equiv 4\int_E^{E_s} dE' \frac{D(E')}{b(E')}$
where $D(E)=D_0 E^{\delta}$ is the diffusion coefficient taken as a power-law in energy. The $e^{\pm}$ energy losses include ICS off the ISRF and the synchrotron emission on the Galactic magnetic field.
The flux of $e^\pm$ at the Earth for a source of age $T$ and distance $d=|{\bf r_{\odot} - r_s}|$ is given by $\Phi_{e^\pm}(E) = \frac{c}{4\pi} \mathcal{N}_e(E,{\bf r=r_{\odot}},t=T).$
We consider as benchmark case the propagation parameters as derived in \cite{dimauro2021multimessenger} from a fit to the latest AMS-02 data for the B/C, antiproton and proton data. 
Energy losses are computed on the interstellar photon populations at different wavelengths following \cite{Vernetto:2016alq}, by taking into account the Klein-Nishina formula for ICS, and on the Galactic magnetic field with intensity $B=3$ $\mu$G. 
As a comparison, we will also implement the {\it SLIM-MED} model derived in 
\cite{genolini2021new}, with the ISRFs taken from \cite{2010A&A...524A..51D} and $B=1$~$\mu$G. 

\section{Simulation and fit strategies}
We produce simulations for the following setups: {\texttt{ModA} (benchmark):} 
 Spin-down and pulsar evolution properties are taken from \texttt{CB20} \cite{Chakraborty:2020lbu}, while the radial surface density of sources is modelled with ~\cite{Lorimer_2006}. $\eta$ and $\gamma_e$ are extracted from uniform distributions, while the propagation in the Galaxy is taking into account following ~\cite{dimauro2021multimessenger}. 
{\texttt{ModB} (radial distribution effect):} 
 Same as \texttt{ModA} but with the radial surface density of sources from ~\cite{2006ApJ...643..332F}. 
{\texttt{ModC} (spin-down properties effect):} 
  Same as \texttt{ModA}, but spin-down properties are taken from \texttt{FK06} \cite{2006ApJ...643..332F}. 
{\texttt{ModD} (propagation effect):} 
  Same as \texttt{ModA} apart for propagation in the Galaxy, modelled as in ~\cite{genolini2021new} (their model {\it SLIM-MED}).
{\texttt{ModE} (kick velocity effect):} 
  Same as \texttt{ModA}, but considering only the $e^{\pm}$ emitted after the escaping of pulsars from the SNR. The birth kick velocities are sampled adopting the distribution \texttt{FK06VB} reported in ~\cite{2006ApJ...643..332F}.
  
For each simulation setup we build and test 1000 simulations. 
 We compute the $e^+$ flux at the Earth as the sum of the primary component due to PWNe emission, and a secondary component due to the fragmentation of CRs on the nuclei of the ISM, taken  from \cite{dimauro2021multimessenger} or \cite{genolini2021new} consistently with the propagation model employed. 
The secondary component enters in our fits with a free normalization factor $A_S$, which we generously let to vary between 0.01 and 3. 
We also let the total flux generated by the sum of all PWNe to be shifted by an overall normalization factor $A_P$. The values of $A_P$ and $A_S$ are obtained for each simulation with the fit procedure.
We fit AMS-02 data \cite{PhysRevLett.122.041102} above 10 GeV, in order to avoid strong influence from solar modulation and other possible low energy effects. Nevertheless, we correct our predictions for solar modulation effects following the force field approximation and leaving the Fisk potential $\phi$ free to vary between 0.4 and 1.2 GV. 
The comparison of our predictions with the AMS-02 $e^+$ data is performed by a standard $\chi^2$ minimization procedure.

\section{Results}

\subsection{Fit to AMS-02 data}
\begin{figure}
\centering
\includegraphics[width=0.45\textwidth]{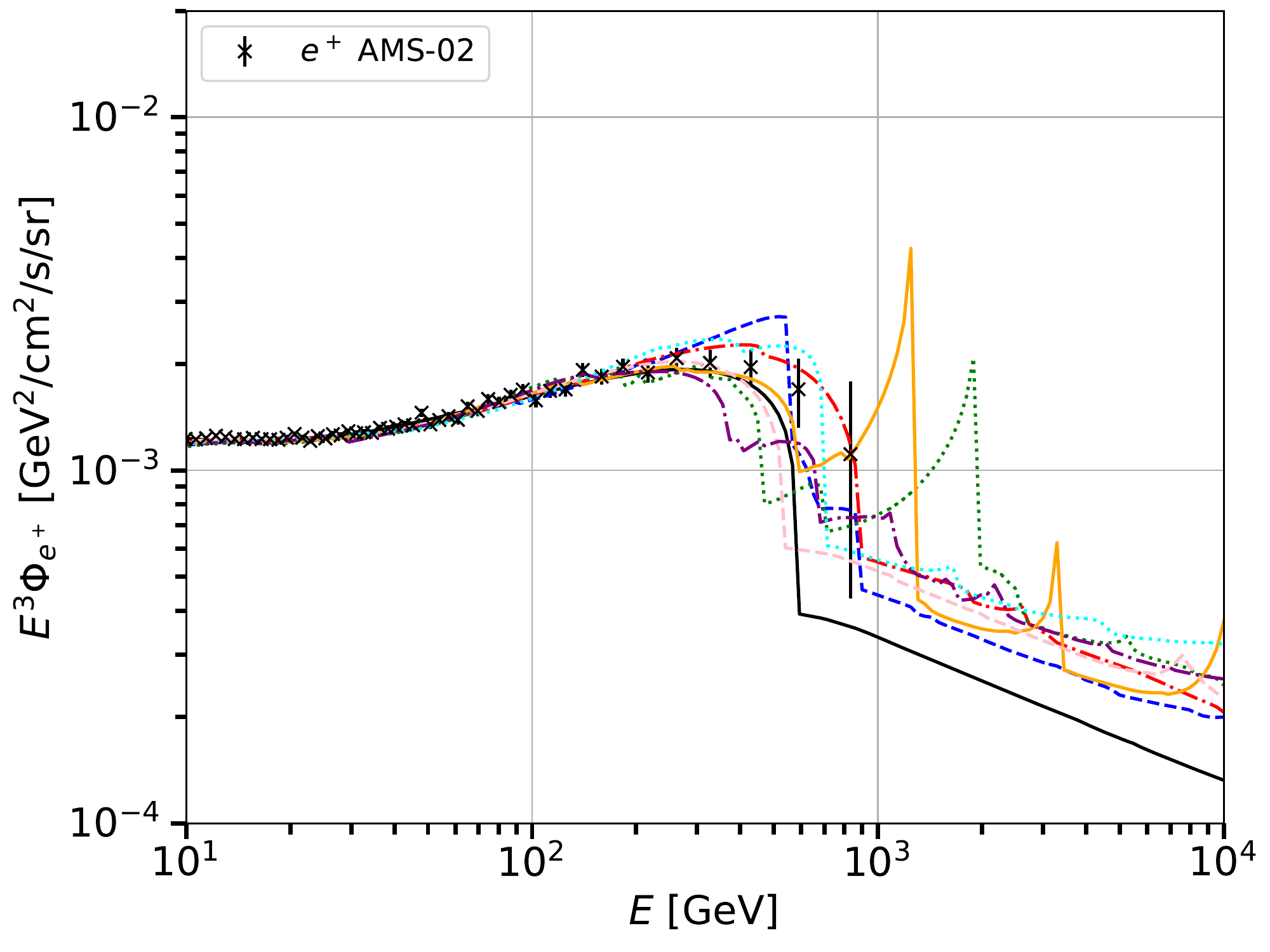}
\includegraphics[width=0.5\textwidth]{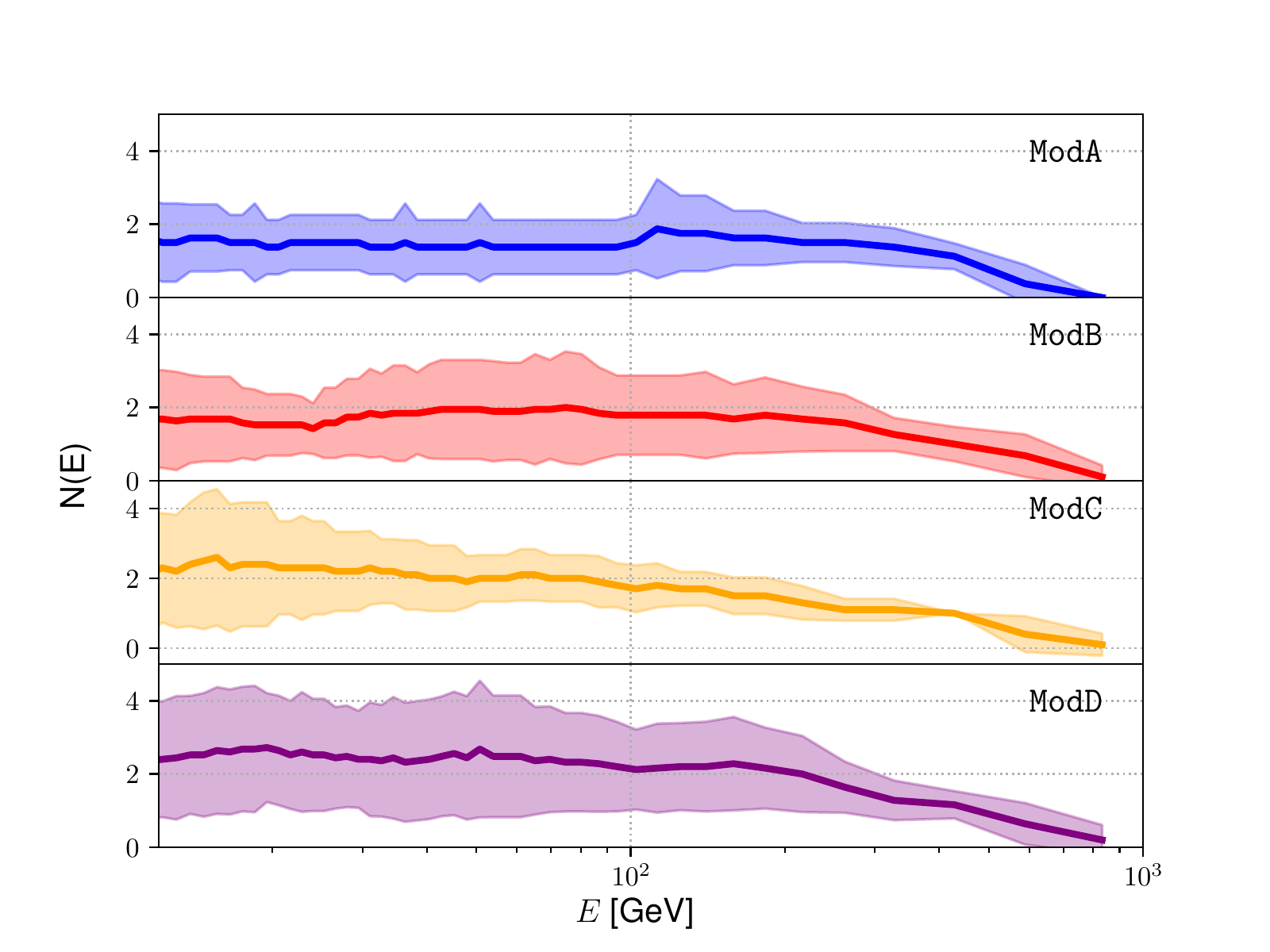}
\caption{Left: Total (secondary plus PWNe) $e^+$ flux obtained from all the 8 simulations within \texttt{ModA} with $\chi^2_{\rm red}<1.5$, along with  AMS-02 data (black points). Right: Mean number of PWNe that satisfy the {\em AMS-02 errors} criterion in the single energy bin of AMS-02 data. We also show the $68\%$ containment band for simulations with $\chi^2_{\rm red}<1.5$.}
\label{Fig R3}
\end{figure}

The fit of the predictions for the total $e^+$ flux to the AMS-02 data is performed for all the 1000 simulations built for each scenario  \texttt{A-B-C-D-E}.
In all the tested setups, the number of mock galaxies with a $\chi^2_{\rm red}<1$ (2) does not exceed 1\% (4\%).
In order to inspect the effects of different simulated Galactic populations, we plot in Figure~\ref{Fig R3}  the total $e^+$ flux for all the pulsar realizations within \texttt{ModA}, and having $\chi^2_{\rm red}<$1.5 on AMS-02 data. 
For energies lower than 200 GeV, differences among the realizations are indistinguishable. The data in this energy range are very constraining. Instead, above around 300 GeV the peculiarities of each galaxy show up, thanks to the larger relative errors in the data.
Above 1 TeV the predictions are unconstrained by data. Nevertheless, all the simulations predict globally 
decreasing fluxes, as expected by energy losses and  continuous $e^\pm$ injection.
However, at these energies  the total flux gets dominated by the secondary component. 

\subsection{How many dominant sources?}
We inspect the average number of sources which contribute the most to the $e^+$ and thus can shape the AMS-02 flux. We adopt two complementary criteria  to estimate the number of sources that are responsible for the most significant contribution of the PWNe $e^+$ emission: (1){\em AMS-02 errors}: we count all the sources that produce a 
    flux higher than the experimental flux error in at least one energy bin above 10 GeV; (2){\em Total flux 1\%}: we count the sources that produce the integral of $\Phi_{e^\pm}(E)$ 
     between 10 and 1000 GeV higher than 1\% of the total integrated $e^+$ flux measured by AMS-02. 
In Figure~\ref{Fig R3} we report the average number of PWNe with the standard deviation (68\% containment band) that contribute in the different energy bins of AMS-02, for configurations with $\chi^2_{\rm red}<1.5$, adopting the {\em AMS-02 errors} criterion.
On average, 2-3 sources shine with a flux at least at the level of AMS-02 $e^+$ data errors.
We also find a decreasing number of dominant sources with increasing energy for all the setup reported. This result is partially induced by the larger experimental errors at high energy, which raise the threshold for the minimum flux that a PWN has to produce in order to satisfy the {\em AMS-02 errors} criterion. 
Moreover, being the age simulated in a uniform interval, the number of young sources responsible for the highest energy fluxes is smaller than for old pulsars, whose $e^+$ have suffered greater radiative cooling. 
Overall, it indicates that only a few sources with a large flux are required in order to produce a good fit to the data.

\subsection{Characteristics of sources dominating the $e^+$ flux}
For each Galactic realization of \texttt{ModA} with $\chi^2_{\rm red}<1.5$, we report in Figure~6a\cite{Orusa:2021tts}  the distance, age and maximum $E^3 \Phi_{e^+}(E)$ of the PWNe satisfying the {\em AMS-02 errors} criterion.
The data require 1 or 2 sources with high maximum $E^3 \Phi_{e^\pm}(E)$, with ages between 400 kyr and 2000 kyr and distances to the Earth less than 3 kpc. These sources produce fluxes peaked between 100 GeV and 500 GeV, allowing good explanation to the data. Fluxes from farther PWNe contribute less to the data. Sources with small maximum $E^3 \Phi_{e^\pm}(E)$ and with ages between 2000 kyr and $10^4$ kyr also satisfy the criterion, with flux peaks below 100 GeV where the secondaries are still the dominant component. We do not find any particular difference between all the simulation setups, 
except for \texttt{ModD} and \texttt{ModE}. Since the {\it SLIM-MED} propagation implemented in \texttt{ModD} produces smoother fluxes, we find also some realizations with few more sources contributing with a bright flux to the $e^+$ data.
In Figure~6b we report the distances, age and maximum $E^3 \Phi_{e^\pm}(E)$ values of the dominant PWNe for the mock galaxies with the worst $\chi^2$. These cases give best-fit to the data with the maximum values allowed by the priors for $A_S$ and the lowest values of $A_P$. 
Moreover, there are not sources which satisfy the {\em AMS-02 errors} criterion with an age between 400 kyr and 2000 kyr. In these galaxies, the trend of $E^3 \Phi_{e^\pm}(E)$ at high energies remains constant or decreases, and does not contribute sufficiently to the data above 50 GeV. 
To compensate this effect, the fit procedure demands the highest value of $A_S$.
We also inspect the distribution of best-fit efficiencies vs initial spin-down energy of each PWN that satisfies the 
{\em AMS-02 errors} criterion, for each simulation with $\chi^2_{\rm red}<1.5$.  The reported efficiencies are obtained
multiplying the simulated $\eta$ values associated to a single source with the $A_P$ obtained from the best fit of the corresponding galaxy. 
The efficiencies have a scattered distribution, and in most cases they have a value between 0.01 and 0.1, confirming the goodness of the $\eta$ interval initially chosen. Data hint at  a slight anti-correlation
between $\eta$ and $W_0$. In order to check that the characteristics of these pulsars are consistent with observations, we compute $\dot{E}$ from $W_0$, finding values quite common in nature. The ATNF catalog \cite{2005AJ....129.1993M} 
lists about 60 sources with $\dot{E}$ values higher than the maximum values obtained from sources in our simulations,  namely $\dot{E}\sim 10^{36}$ erg s$^{-1}$. We do not directly compare the $W_0$ values, since for the sources of the ATNF catalog to compute $W_0$ we need to assume arbitrarily the value of $n$ and $\tau_0$. 
Instead, in our simulations we sample $n$ and we compute $\tau_0$ from the simulated parameters like $P_0$, that is also strictly connected to $W_0$.

\section{Summary \& Conclusions}

Precise AMS-02 $e^+$ data give us unprecedented insights on primary CR antimatter sources, but raise also new questions on the CR production and transport in our Galaxy.
In this contribution, the  AMS-02 $e^+$ data are used to constrain the main properties of the Galactic pulsar population and of the PWN acceleration needed to explain the observed CR flux. To this aim, we simulate a large number of  Galactic pulsar populations, calibrated on ATNF catalog observations. The result is fitted to AMS-02 $e^+$ data. 
Our simulations are conducted under different hypothesis about the pulsar spin-down and evolution properties, their radial distribution, and changing the propagation models for $e^\pm$ propagation in the Galaxy, following the most recent self-consistent modelings available in the literature.

Independently of the simulation scenario, we find that the vast majority of the galaxies realizations produce several wiggles in the total contribution and therefore they do not fit well the data. In all the tested setups, the number of mock galaxies with a $\chi^2_{\rm red}<1$ (2) does not exceed 1\% (4\%). 
The different features of the flux from single PWNs are due to the peculiar combination of the input parameters. We notice that the secondary flux, while decreasing with energy, practically forbids the realization of sharp cut-offs in the $e^+$ spectrum above TeV energies.
The galaxy realizations that fit properly the AMS-02 $e^+$ data have between 2-3 sources that produce a $e^+$ yield at the level of the data errors. Moreover, these pulsars provide a smooth spectrum that cover a wide energy range.
We find that the dominant contribution comes from sources located  between 1 and 3 kpc from the Earth. 
Despite the smaller effect from radiative cooling, the flux from sources within 1 kpc is lower due to the paucity of sources set along the spiral arms of the Galaxy. 
Sources dominating the observed spectrum have ages between 400 kyr and 2000 kyr and distances to the Earth less than 3 kpc. These sources produce fluxes peaked  between 100 GeV and 300 GeV, where AMS--02 data are the most constraining. 
Finally, we do not find any particular distribution for the pulsar efficiencies. In most cases they have a value between 0.01 and 0.1 consistently with what found in our previous papers \cite{DiMauro:2019yvh}.

Our results are compared with other recent attempts of simulating pulsar populations \cite{Cholis_2018,Fornieri_2020,Evoli_2021} in the last section of \cite{Orusa:2021tts}.
To move forward, we anticipate refinements on the acceleration, release and propagation models around pulsars using multiwavelength probes, as well as a detailed analysis of pulsar in source catalogs possibly matching the characteristics suggested by the simulations.

\section*{References}

\bibliography{manconi}

\end{document}